# Electric-field modification of interfacial spin-orbit field-vector

L. Chen[1], M. Gmitra[2], M. Vogel[1], R. Islinger[1], M. Kronseder[1], D. Schuh[1], D. Bougeard[1], J. Fabian[2], D. Weiss[1] and C. H. Back[1]

[1]Institute of Experimental and Applied Physics, University of Regensburg, 93040 Regensburg, Germany.

[2]Institute of Theoretical Physics, University of Regensburg, 93040 Regensburg, Germany.

Correspondence and requests for materials should be addressed to C.H.B. (e-mail: christian.back@ur.de).

**Current induced spin-orbit magnetic fields (iSOFs), arising either in single-crystalline ferromagnets with broken inversion symmetry[1,2] or in non-magnetic metal/ferromagnetic metal bilayers[3,4], can produce spin-orbit torques which act on a ferromagnet's magnetization, thus offering an efficient way for its manipulation. To further reduce power consumption in spin-orbit torque devices, it is highly desirable to control iSOFs by the field-effect, where power consumption is determined by charging/discharging a capacitor[5,6]. In particular, efficient electric-field control of iSOFs acting on ferromagnetic metals is of vital**

**importance for practical applications. It is known that in single crystalline Fe/GaAs (001) heterostructures with $C_{2v}$ symmetry, interfacial SOFs emerge at the Fe/GaAs (001) interface due to the lack of inversion symmetry[7,8]. Here, we show that by applying a gate-voltage across the Fe/GaAs interface, interfacial SOFs acting on Fe can be robustly modulated via the change of the magnitude of the interfacial spin-orbit interaction. Our results show that, for the first time, the electric-field in a Schottky barrier is capable of modifying SOFs, which can be exploited for the development of low-power-consumption spin-orbit torque devices.**

Experiments in non-magnetic metal (NM)/ferromagnetic metal (FM) hetero-structures, have shown that current induced interfacial non-equilibrium spin polarization due to spin-orbit interaction, originating either from the Rashba effect at the NM/FM interface[3] or the spin Hall effect in the bulk of NM[4], can be used to reverse the magnetization direction of ferromagnets, thus paving the way for the developement of low-power consumption magnetic memory devices. It is convenient to parameterize the current-induced non-equilibrium spin polarization as effective spin-orbit magnetic field (iSOFs)[1,9], where the magnitude of iSOFs is proportional to the applied current density in the NM[10]. To efficiently reorientate the magnetization using low currents, materials with large charge-to-spin conversion efficiency have been searched for and remarkable progress has been made in the past few years[11-16]. An alternative route towards efficient manipulation of magnetization is electric-field control, e.g., using a capacitor structure where the energy consumption is, in principle, determined by charging/discharging the capacitor[6].

Consequently, to further reduce power consumption in spin-torque devices, field-effect control of the iSOFs is highly desirable. Recently, it has been shown that electric-field control of current-induced effective magnetic fields can be realised in a magnetic topological insulator via a top gate by which the Fermi level within the Dirac bands gets shifted[17]. However, research on electric-field control of iSOFs acting on ferromagnetic metals, i.e., in NM/FM bi-layers, is just at its beginning[18]. Here, the experimental difficulty is that spin-orbit interaction at NM/FM interfaces can not be tuned effectively as the short screening length in metals shields external electric-fields[18,19].

Besides NM/FM bilayer systems, also single crystalline ferromagnets lacking space-inversion symmetry, feature iSOFs which can induce spin-orbit torques acting on the ferromagnet itself[1,2]. Recent examples are bulk (Ga,Mn)As[1,2,20,21], NiMnSb[22] and Fe/GaAs (001)[8]. For (Ga,Mn)As and NiMnSb, the iSOFs occur mainly in the bulk resulting from the ferromagnets' bulk inversion asymmetry; while for Fe/GaAs (001), the iSOFs have an interfacial origin[7], thus enabling electric-field control. Here, we report electric-field control of iSOFs at the Fe/GaAs (001) interface. By applying a gate-voltage across the interface, the interfacial spin-orbit field-vector acting on Fe can be varied.

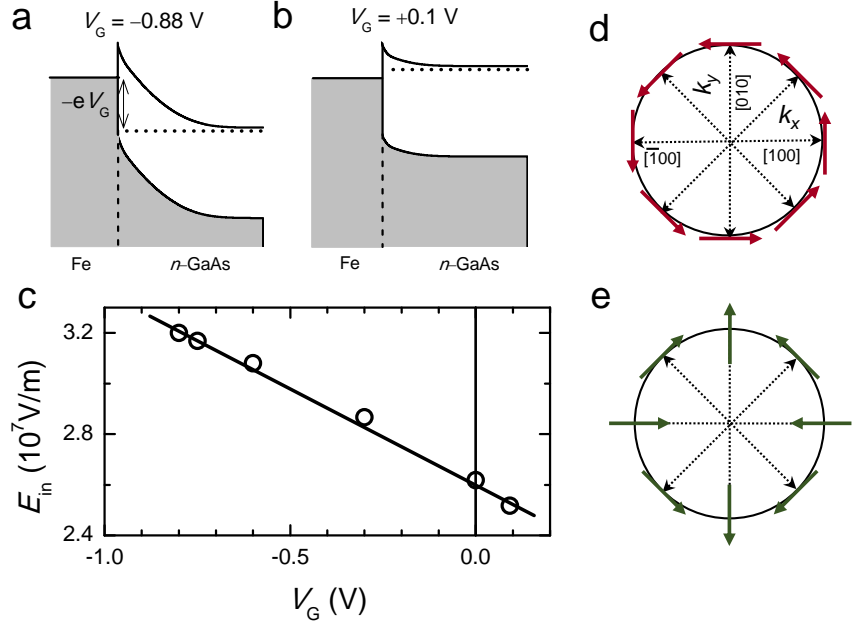

**Figure 1 | Electric-field modulation of spin-orbit fields at the Fe/GaAs (001) interface.** Schematic band bending of a Fe/*n*-GaAs Schottky diode under reverse (**a**) and forward (**b**) bias. (**c**) Calculated interfacial electric-field $\mathbf{E}_{in}$ obtained by solving the one-dimensional Poisson equation for the non-constant doping profile of *n*-GaAs (Methods). The magnitude of interfacial spin-orbit interaction is proportional to $\mathbf{E}_{in}$. Wave-vector **k** dependence of Rashba- (**d**) and Dresselhaus (**e**) spin-orbit fields which dominate at the epitaxial Fe/GaAs interface with $C_{2v}$ symmetry.

The mechanism of the electric-field modulation is shown in Fig. 1. Due to the $C_{2v}$ symmetry of the interface, Bychkov-Rashba- and Dresselhaus-like spin-orbit interaction emerges at the Fe/GaAs (001) interface, and the corresponding **k**-space dependence of the equilibrium spin-orbit magnetic fields, SOFs, $\mathbf{h}_{eff}$ can be written as[23]

$$\mu_0 \mathbf{h}_{eff} = \frac{1}{\mu_B}\left(-\beta k_x - \alpha k_y, \alpha k_x + \beta k_y\right),\qquad(1)$$

where $\mu_0$ is the magnetic constant, $\mu_B$ is Bohr's magneton, $k_x$ ($k_y$) is the wave-vector **k** along the [100] ([010]) direction, and $\beta$ ($\alpha$) characterizes the strength of the Dresselhaus (Bychkov-Rashba) spin-orbit coupling. The magnitude of $\beta$ and $\alpha$ depends linearly on the interfacial electric-field $E_{in}$,

$$\beta = C_\beta E_{in} \text{ and } \alpha = C_\alpha E_{in},\qquad(2)$$

where $C_\beta$ ($C_\alpha$) is the coefficient of Dresselhaus (Bychkov-Rashba) spin-orbit interaction[24]. The sample investigated here is a Fe/n-GaAs heterostructure, in which $\mathbf{E}_{in}$ consists of the internal built-in electric-field (pointing from GaAs to Fe due to the larger work function of n-GaAs[25]) and the external electric-field with its direction depending on the polarity of gate-voltage $V_G$. The advantage of the Fe/n-GaAs Schottky tunnel junction is that the depleted n-GaAs below the Fe layer serves as a dielectric insulator across which an external electric-field can be applied (Figs. 1a and b), e.g., biasing in reverse (forward) direction increases (decreases) $E_{in}$. Thus, by tuning the gate-voltage, both strength and direction of the SOFs can be tuned. The calculated $V_G$ dependence of $E_{in}$ of the device layout is presented in Fig. 1c (Methods), which shows an essentially linear behavior within the investigated $V_G$ range.

Below we demonstrate electric-field tuning of iSOFs in epitaxially grown Fe/n-GaAs samples with different Fe thickness and n-GaAs doping concentrations. All samples show consistent results and we focus here on data taken from the sample with with respect to the [100] orientation of GaAs. (**b**) d.c. voltage spectra for several gate-voltages for a [100]-oriented device measured at a microwave frequency of 12.5 GHz and $\varphi_M = 225°$. The input microwave power is 10 mW. An offset voltage, $V_{offset}$, has been subtracted. (**c**) Typical Lorentzian fit of the voltage spectrum, from which the magnitudes of $V_{sym}$ and $V_{a\text{-}sym}$ are obtained. (**d**) $V_G$ dependence of $V_{a\text{-}sym}$ and $V_{sym}$ for $\varphi_M = 225°$. Magnetization angle $\varphi_M$ dependence of $V_{a\text{-}sym}$ (**e**) and $V_{sym}$ (**f**) under different $V_G$ for a [100]-oriented device. The solid lines in (**e**) and (**f**) are fits to Eqs. 3 and 4, respectively.

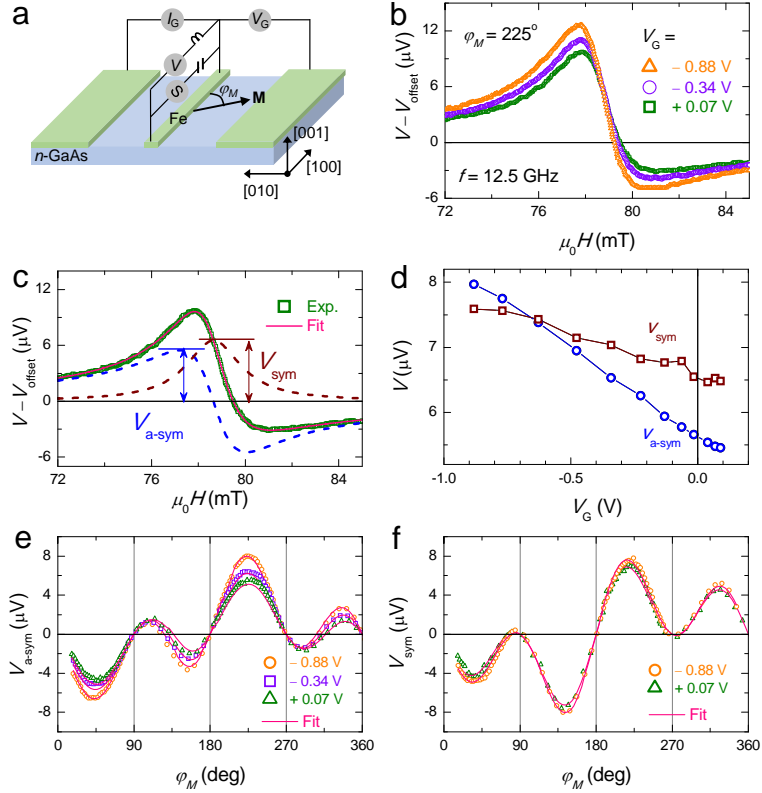

**Figure 2 | Electric-field modulation of induced spin-orbit fields measured by spin-orbit ferromagnetic resonance.** (**a**) Schematic of the circuit and device structure for spin-orbit ferromagnetic resonance measurements. A microwave current **j**($t$), generated by a signal generator, passes through a bias tee into the Fe/GaAs (001) layer, where time dependent interfacial iSOFs **h**($t$) are generated due to the spin-orbit interaction at the Fe/GaAs interface. **h**($t$) drives the magnetization dynamics **M**($t$), and leads to a resistance variation $R(t)$ due to the anisotropic magneto-resistance of Fe. The coupling between $R(t)$ and $j(t)$ produces a rectified d.c. voltage $V$. By measuring $V$, the magnitude of the interfacial iSOFs can be quantified. To modulate the iSOFs by an electric-field, Fe/$n$-GaAs samples have been used, where a Schottky barrier is formed at the interface. Thus, by adding a gate-voltage $V_G$ between Fe and $n$-GaAs, the magnitude of the iSOFs can be controlled. Here, $\varphi_M$ is the magnetization angle

4-nm Fe thickness. Fe stripes with dimension of 10 μm × 200 μm oriented along four main crystal directions, i.e., [100], [110], [010], and [$\bar{1}$10], which determine the prevailing **k**-vector directions, are fabricated by electron-beam lithography (for example, see Fig. 2a for a [100]-orientated device and Methods). To quantify the magnitude of the interfacial iSOFs, we use the spin-orbit ferromagnetic resonance technique[8,20]. Microwaves with a power of 10 mW and a frequency of 12.5 GHz are applied to the Fe stripe through a bias tee. The rectified d.c. voltage spectrum induced by the anisotropic magneto-resistance effect of Fe is measured by sweeping an external magnetic field $H$ across the ferromagnetic resonance (FMR) of Fe. For measuring the gate-voltage dependence of the d.c. voltage, a gate-voltage $V_G$ is applied across the Fe/$n$-GaAs interface (Methods). Since the detection direction of the d.c. voltage is parallel to the microwave current, the dynamic tunneling anisotropic magneto-resistance effect[26] and the dynamic Hanle effect[27], only detectable transverse to the microwave current, cannot contribute to the signal.

Figure 2b shows the d.c. voltage spectra as a function of $V_G$ at a magnetization angle $\varphi_M$ of 225° for the [100]-oriented device measured at room temperature. The modulation of the spectra with $V_G$ is evident. As shown in Fig. 2c, each spectrum can be well fitted by combining a symmetric ($L_{sym} = \Delta H^2 / [4(H-H_R)^2 + \Delta H^2]$) and an anti-symmetric Lorentzian ($L_{a-sym} = -4\Delta H(H-H_R) / [4(H-H_R)^2 + \Delta H^2]$), $V-V_{offset} = V_{sym}L_{sym} + V_{a-sym}L_{a-sym}$, where $H_R$ is $H$ at FMR, $\Delta H$ is the full width at half maximum, $V_{offset}$ is the offset voltage, and $V_{sym}$ ($V_{a-sym}$) is the magnitude of the symmetric (anti-symmetric) component of the d.c. voltage. The fitting procedure gives values for $H_R$, $\Delta H$, $V_{sym}$ and $V_{a-sym}$. The results show that both $H_R$ and $\Delta H$ do not depend on $V_G$, indicating that the bulk magnetic properties of Fe are not influenced by the external electric-field (see Supplementary Information). However, a clear modification of $V_{a-sym}$ and $V_{sym}$ is observed in Fig. 2d. The dependence of $V_{a-sym}$ and $V_{sym}$ on $V_G$ shows distinctly different slopes, indicating different origins for $V_{a-sym}$ and $V_{sym}$ (see Supplementary Information).

Figure 2e shows the dependence of $V_{a-sym}$ on magnetization direction $\varphi_M$ at $V_G$ of −0.88 V, −0.34 V and +0.07 V for the [100]-oriented device. A clear modulation of $V_{a-sym}$ by $V_G$ can be seen. By analyzing the angular dependence, the magnitude of the in-plane iSOFs at each $V_G$ can be extracted using

$$V_{\text{a-sym}}^{[100]} = -\frac{\Delta \rho j_{\text{Fe}} l}{2M} \text{Re}(\chi^{\text{I}})\left(-h_{\text{D}}^{\text{I}} \sin\phi_M + h_{\text{R}}^{\text{I}} \cos\phi_M\right)\sin 2\phi_M, \quad (3)$$

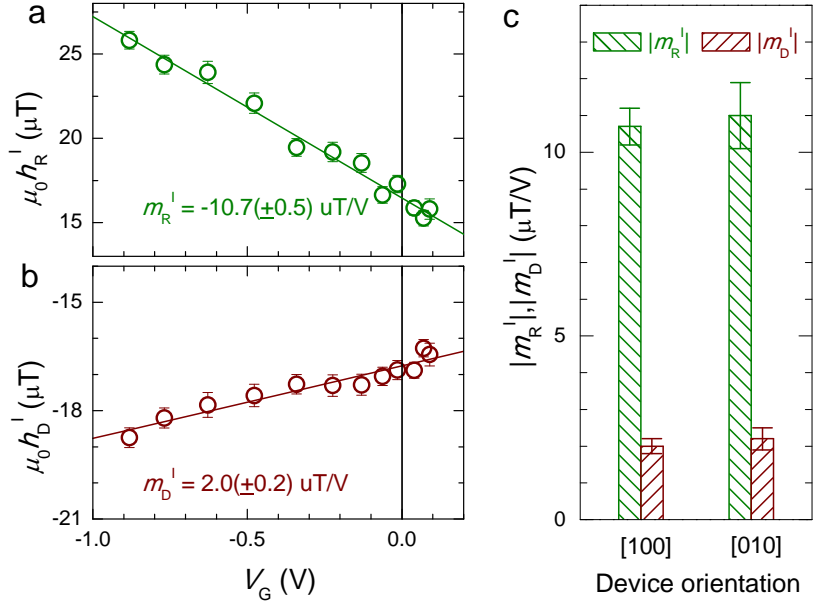

**Figure 3 | Gate-voltage dependence of in-plane induced spin-orbit fields.** $V_G$ dependence of in-plane Bychkov-Rashba- (**a**) and Dresselhaus- (**b**) iSOFs obtained from the [100]-oriented device. The solid lines are linear fits, and the modification rate of Rashba-iSOF $m_R^I$ and Dresselhaus-iSOF $m_D^I$ is obtained from the slopes. (**c**) Absolute value of $m_R^I$ and $m_D^I$ for devices along [100] and [010] directions. The magnitude of $|m_R|$ is larger than $|m_D|$ for both orientations. The error bars in (**c**) are standard deviation from the linear fits.

where $h_D^I$ ($h_R^I$) is the current induced in-plane iSOF due to Dresselhaus (Bychkov-Rashba) spin-orbit interaction, $\Delta\rho$ is the magnitude of the anisotropic magneto-resistance of Fe, $j_{Fe}$ is the microwave current density in Fe, $l$ is the length of the device, $M$ is the magnetization, and $\text{Re}(\chi^I)$ is the real part of the diagonal component of the magnetic susceptibility due to in-plane excitation[8]. The magnitude of $\Delta\rho$, $j_{Fe}$, $M$ and Re($\chi^I$) is confirmed to be $V_G$ independent (see Supplementary Information). Thus, by using constant values of $\mu_0 M$ = 2.1 T, $\Delta\rho$ = 1.1×10$^{-9}$ $\Omega$m, $l$ = 200 μm and $j_{Fe}$ = 1.0×10$^{10}$ Am$^{-2}$, the angular dependence of $V_{a\text{-sym}}$ at each $V_G$ can be well reproduced by Eq. (3) using $h_D^I$ and $h_R^I$ as adjustable parameters. The extracted values of $h_R^I$ and $h_D^I$ are shown as a function of $V_G$ in Figs. 3a and b, respectively. One can see that the absolute value of $h_R^I$ and $h_D^I$ increases (decreases) at negative (positive) $V_G$. Besides the polarity, the dependence of $h_R^I$ ($h_D^I$) on $V_G$ is basically linear, which is expected from the calculated linear dependence of the interfacial electric-field on $V_G$ (see Fig. 1c and Ref. 28). The modification rate can be quantified by the slope $m$ extracted from the linear fits (solid lines). As shown in Fig. 3c, for both [100] and [010] orientations, $|m_R^I|$ is ~ 5 times larger than $|m_D^I|$, indicating that the modulation rate of the in-plane iSOFs originating from the Bychkov-Rashba spin-orbit interaction is larger than that from the Dresselhaus one. The different amplitudes of $|m_R^I|$ and $|m_D^I|$ also exclude the possibility that the observed modulation is an artifact due to changes of $j_{Fe}$ with $V_G$. If the modulation were dominated by a $V_G$ dependent $j_{Fe}$, $|m_R^I|$ would be equal to $|m_D^I|$ according to Eq. 3 (see Supplementary Information). We also show in the Supplementary Information that other possible excitations, e.g., the Oersted field as well as the bulk Dresselhaus field induced by the inductive microwave current in un-depleted $n$-GaAs, are negligibly small.

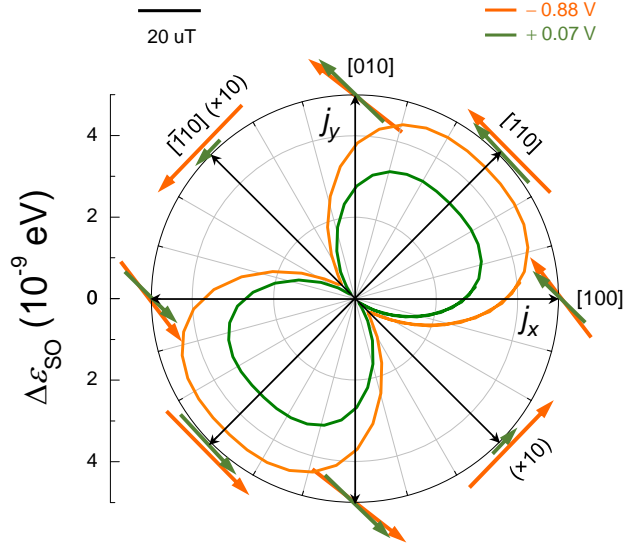

**Figure 4 | Tuning of in-plane induced spin-orbit fields.** Polar plot of in-plane iSOFs. The arrows represent direction and relative strength of $\mathbf{h}_{eff}$, $\mathbf{h}_{eff} = \mathbf{h}_R^I + \mathbf{h}_D^I$, along the four main directions for a current density of $1\times10^{10}$ Am$^{-2}$. Note that the magnitude of the iSOFs for [$\bar{1}10$] has been amplified by a factor of 10, and the orange arrows along [110] and [$\bar{1}10$] orientations have been shifted for clarity. The solid lines show the polar plot of the spin-orbit energy splitting, $\Delta\varepsilon_{SO} = 2\mu_B|\mu_0\mathbf{h}_{eff}|$ for $V_G = -0.88$ V and $+0.07$ V. The full angular dependence of $|\mathbf{h}_{eff}|$ is extrapolated from the magnitude of field-vectors along the four main orientations.

The complete angular dependence of the in-plane iSOFs, which is obtained from devices oriented along the four main crystal directions, is shown as a polar plot in Fig. 4 for $V_G = -0.88$ V and $+0.07$ V. In the plot, the lengths of the arrows represent the magnitude ($|\mathbf{h}_{eff}| = |\mathbf{h}_R^I + \mathbf{h}_D^I|$), and the angles between the arrows and current directions represent the direction of the iSOFs. For the [100] ([010]) direction, both the length and angle of the arrows change with $V_G$; in contrast to the [110] and [$\bar{1}10$] directions, where only the length of the arrows is modified by $V_G$ as expected from the **k**-space alignment of Bychkov-Rashba and Dresselhaus SOFs (see Figs. 1d and e). These results clearly demonstrate an electric-field modulation of the spin-orbit field-vectors. Note that the length of the arrow along [110] is much larger than that along [$\bar{1}$10] (the amplitude for [$\bar{1}$10] has been amplified by a factor of 10), which is due to the fact that Dresselhaus and Bychkov-Rashba fields add along [110] but get subtracted along [$\bar{1}$10] (see Figs. 1d and e). The dramatic amplitude difference for [110] and [$\bar{1}$10] directions also excludes the Oersted field in Fe as a possible driving force for the magnetization dynamics. Moreover, as shown by the solid lines in Fig. 4, the modulation effect becomes much clearer from the polar plot of the spin-orbit energy splitting, $\Delta\varepsilon_{SO} = 2\mu_B|\mu_0\mathbf{h}_{eff}|$[23]. Here, the full angular dependence of $|\mathbf{h}_{eff}|$ is extrapolated from the magnitude of the magnetic field-vectors obtained from devices for four main orientations.

Besides the in-plane iSOFs, the presence of $V_{sym}$ also implies the existence of a sizeable out-of-plane iSOF, $h^z$, which is due to the combined effects of in-plane spin polarization and exchange coupling[8,21,29]. The out-of-plane iSOFs act on the magnetization and produce an anti-damping like torque. Similar out-of-plane spin polarization has also been observed in nonmagnetic semiconductor structures caused by the joint effect of in-plane spin polarization and Zeeman coupling through an external magnetic field[30-32]. However, the coupling (Zeeman field ~ $10^{-5}$ eV) in semiconductors is much smaller than that in ferromagnets with reduced symmetry (exchange field ~ 30 meV for GaMnAs and ~ 1 eV for Fe). The magnitude of $h^z$ is magnetization direction dependent[8,21]. Taking the [100]-oriented device as an example, the dependence can be expressed as, $h^z = h_R{}^z \cos\varphi_M - h_D{}^z \sin\varphi_M$, where $h_R{}^z$ ($h_D{}^z$) originates from Bychkov-Rashba (Dresselhaus) spin-orbit interaction[8]. Thus, the $\varphi_M$ dependence of $V_{sym}$ shown in Fig. 2f can be described by

$$V_{\text{sym}}^{[100]} = \frac{\Delta\rho\, j_{Fe}\, l}{2M} \operatorname{Im}\left(\chi_{\mathbf{a}}^{\mathbf{o}}\right)\left(h_{\mathbf{R}}^{\mathbf{z}} \cos\phi_M - h_{\mathbf{D}}^{\mathbf{z}} \sin\phi_M\right)\sin 2\phi_M , \qquad (4)$$

where Im($\chi_a^O$) is the imaginary part of the off-diagonal component of the magnetic susceptibility due to the out-of-plane excitation[8]. The obtained gate-voltage dependence of $h_R^z$ and $h_D^z$ is summarized in Figs. 5a and b, respectively. Similar to the in-plane case, a clear modulation can be seen and the modulation amplitude of $h_R^z$ is larger than $h_D^z$, indicating again a stronger modulation of the Bychkov-Rashba spin-orbit interaction.

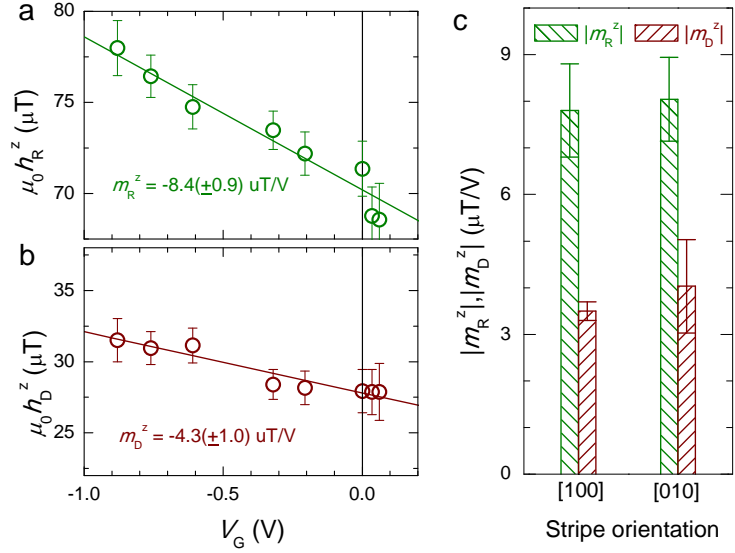

**Figure 5 | Gate-voltage dependence of the out-of-plane induced spin-orbit fields.**
(**a**) Gate-voltage $V_G$ dependence of the out-of-plane induced spin-orbit fields originating from Bychkov-Rashba (**a**) and Dresselhaus (**b**) spin-orbit interaction for the [100]-oriented device. The solid lines in (**a**) and (**b**) are linear fits to the experimental data. From the slope, the modification rate of $h_R{}^z$ and $h_D{}^z$ is obtained. (**c**) Absolute values of $m_R{}^z$ and $m_D{}^z$ obtained from the linear fits of the out-of-plane iSOFs for devices along [100] and [010] directions. The error bars in (**c**) are standard deviation from the linear fits.

Now, we discuss the mechanism for the electric-field modulation of iSOFs. According to Eqs. 1 and 2, the modulation is directly related to the modulation of the strength of the spin-orbit interaction. This modulation can be in general different for $h_R$ and $h_D$, as the two fields originate from two inequivalent spatial crystal orientations [110] and [$\bar{1}$10] at the Fe/GaAs interface[7]. We indeed see that the electric-field modification of the Bychkov-Rashba and Dresselhaus field is different, see Figs. 3c and 5c. The smaller $|m_D|$ is consistent with previous measurements in semiconductor quantum wells[33-35], where the magnitude of Dresselhaus spin-orbit interaction is less sensitive to the external electric-field compared to Bychkov-Rashba spin-orbit interaction. The different gate dependences for in-plane and out-of-plane components of both Bychkov-Rashba- and Dresselhaus-SOFs have another origin. The in-plane iSOFs are formed by spin accumulation $\sigma_{//}$ at the Fermi level[36], while, the out-of-plane iSOFs originate from the whole bands, known as intrinsic effect[8,21,29]. Another possible origin of $h^z$ could be the spin-transfer torque effect due to the absorption of in-plane spin accumulation[37,38], i.e., $\mathbf{h}^z \sim \mathbf{M} \times \mathbf{\sigma}_{//}$, as in case of NM/FM bi-layers. If $\mathbf{h}^z$ originates dominantly from spin-transfer torque, applying negative gate-voltage can only decrease the absorption and thus $h^z$. Since we observe a clear increase of $h^z$ at negative bias, spin-transfer torque effects can also be excluded.

The magnitudes of $m_R$ and $m_D$ are sizeable. Density function theory (DFT) results have predicted an efficient control of interfacial spin-orbit field for transverse electric-fields of the order of 1 V/nm (see supplementary material of Ref. 7). Our electric-fields are much smaller, about 0.01 V/nm. In DFT calculations thin slabs of Fe/GaAs without doping were studied, while our structure comprises a wide (~100 nm) Schottky barrier. In thin slabs large electric-fields are needed to modify the interfacial orbital hybridization between Fe and GaAs atoms to change interfacial spin-orbit field. In our samples we believe that we have interfacial states that penetrate deep enough into GaAs to get influenced by the Schottky barrier. Applying the gate-voltage then changes the barrier which, in turn, changes the overlap of the interfacial electron states with Fe. The modification of SOFs follows, since they are determined by the interfacial overlap. The assumption, that we have an additional conducting interfacial channel, that leads effectively to spin accumulation, is supported by the sizeable interfacial resistance at reverse gate-voltage in the junctions ($\sim$ 0.19 k$\Omega$/µm$^2$ for $V_G$ = −0.5 V, see Fig. S1b).

Finally, we want to mention that in conventional approaches of electrically controlling iSOFs acting on a ferromagnetic metal[18], a top-gate field-effect transistor has been utilized to change the carrier population of the magnetic material. As a consequence, effective control of the iSOFs acting on the ferromagnetic metal requires that the thickness of the ferromagnet should be ultra-thin (usually below 1 nm) due to the short screening length of a metal. However, the method presented here, i.e., voltage control by changing the magnitude of the interfacial spin-orbit interaction in a Schottky diode, is distinctly different from the conventional approach, and not limited by the thickness of the ferromagnetic metal (4 nm in this study). This unique approach thus provides an alternative way of electric-field control. The efficient electric-field control suggests that an electric-field in a Schottky diode can be used for the development of low-power consumption spin-orbit torque devices for magnetization manipulation.

**Methods:**

**Calculation of interfacial electric-field.** To calculate the interfacial electric-field $E_{in}$ for Fe/$n$-GaAs, we solve the one-dimensional Poisson equation, $d^2\varphi/dx^2 = -\rho/\varepsilon_r\varepsilon_0$, where $\varphi$ is electric potential, $\rho$ is charge density, $\varepsilon_r$ is the relative permittivity, and $\varepsilon_0$ is the permittivity in vacuum. In the calculation, the work function of Fe is set to 4.7 eV, and the electron affinity of GaAs is 4.15 eV. Since the experiment is carried out at room temperature, we consider all the donors of $n$-GaAs to be ionized, thus the donor concentration $N_D$ is equal to the carrier concentration $n$. The 15 nm graded $n \rightarrow n+$ junction is discretized into 5 steps, and the Fermi-Dirac statistics has been employed. Having obtained the distribution of $\varphi$ on $x$, the interfacial electric-field at each gate-voltage can be determined from $E_{in} = -d\varphi/dx|_{interface}$.

**Sample preparation.** The Fe/$n$-GaAs heterostructure utilized in this experiment is grown on semi-insulating GaAs (001) substrates in a molecular-beam epitaxy cluster.

The semiconductor part is grown in a III-V MBE, consisting of the following layers: a 100-nm undoped GaAs buffer layer, 250 nm Si-doped $n$-GaAs layer with carrier concentration of $4\times10^{16}$ cm$^{-3}$, a 15 nm $n \to n^+$ transition layer, and a 10 nm $n^+$-GaAs layer with carrier concentration of $1\times10^{18}$ cm$^{-3}$. The thickness and carrier density of the $n$-GaAs layers have been designed to minimize the inductive microwave current density in non-depleted GaAs, which could drive the magnetization dynamics through Oersted field and bulk-Dresselhaus field (see Supplementary Information). After that, the wafer is transferred to a metal MBE without breaking the vacuum, and 4-nm-thick Fe is deposited epitaxially. To avoid Fe oxidation, 2.5-nm thick Al capping layer is also deposited in the same chamber, which is estimated to be fully oxidized in air. The growth temperature for Fe is 75 ºC.

**Device.** First, a Fe/$n$-GaAs mesa with dimension of 200 μm × 900 μm is defined by combining Ar ion-beam milling and chemical wet etching (CH$_3$COOH:H$_2$O:H$_2$O$_2$ = 5:5:1) down to the insulating substrate. Then, SO-FMR Fe-stripes with dimensions of 200 μm × 10 μm are defined on the mesa and the residual Fe in the $n$-GaAs channel region is etched away. We also use Fe/$n$-GaAs with a large area of 400 μm × 300 μm as remote contacts (Fig. 2a). Finally, Ti/Au electrodes and bonding pads are fabricated from 15 nm Ti and 300 nm Au by evaporation. Devices with different orientations have been fabricated to measure the symmetry of iSOFs under electric-field modulation.

**Measurements.** As shown in Fig. 2a, microwave currents generated by the signal generator pass into the Fe stripe and drive the magnetization dynamics due to current induced SOFs at the Fe/GaAs interface. The d.c. voltage is measured across the stripe due to the coupling of the microwave current and the magnetization dynamics through anisotropic magnetoresistance effect of Fe. A bias tee is used to separate the d.c. voltage from the microwave background. The microwave frequency is 12.5 GHz, and the microwave power is set as low as possible (10 mW in this experiment). Under this condition the $I_G$-$V_G$ curve and thus the Schottky barrier remains unaffected by microwaves (see Supplementary Information). To modulate the SOF by external electric-fields, a bias current $I_G$ with different polarity is applied between the Fe stripe and the remote contact, which can be converted to a gate-voltage by the interfacial $I_G$-$V_G$ characteristic (see Supplementary Information). The measurements are performed between −0.88 V and +0.07 V. Since the d.c. bias-current (~ 2 µA) produces a background voltage in the millivolt range, the d.c. voltage is detected by a lock-in amplifier by modulating the microwaves on and off at 888 Hz using a function generator. All measurements are performed at room temperature.

**References**

1. Chernyshov, A. *et al*. Evidence for reversible control of magnetization in a ferromagnetic material by means of spin-orbit magnetic field. *Nature Phys*. **5**, 656-659 (2009).

2. Endo, M. *et al*. Current induced effective magnetic field and magnetization reversal in uniaxial anisotropy (Ga,Mn)As. *Appl. Phys. Lett.* **97**, 222501 (2010).

3. Miron, I. M. *et al*. Perpendicular switching of a single ferromagnetic layer induced by in-plane current injection. *Nature* **476**, 189-193 (2011).

4. Liu, L. Q. *et al*. Spin-torque switching with the giant spin Hall effect of Tantalum. *Science* **336**, 555-558 (2012).

5. Ohno, H. Electric-field control of ferromagnetism. *Nature* **408**, 944-946 (2000).

6. Matsukura, F. *et al*. Control of magnetism by electric fields. *Nature Nanotech*. **10**, 209-220 (2015).

7. Gmitra, M. *et al*. Magnetic control of spin-orbit fields: A first-principle study of Fe/GaAs junctions. *Phys. Rev. Lett*. **111**, 036603 (2013).

8. Chen, L. *et al*. Robust spin-orbit torque and spin-galvanic effect at the Fe/GaAs (001) interface at room temperature. *Nature Commun.* **7**, 13802 (2016).

9. Kim, J. *et al*. Layer thickness dependence of the current-induced effective field vector in Ta | CoFeB | MgO. *Nature Mater*. **12,** 240-245 (2013).

10. Garello, K. *et al*. Symmetry and magnitude of spin-orbit torques in ferromagnetic heterostructures. *Nature Nanotech*. **8**, 587-593 (2013).

11. Soumyanarayanan, A. *et al*. Emergent phenomena induced by spin-orbit coupling at surfaces and interfaces. *Nature* **539**, 509-517 (2016).

12. Mellnik, A. R. *et al*. Spin-transfer torque generated by a topological insulator. *Nature* **511**, 449-451 (2014).

13. Fan, Y. B. *et al*. Magnetization switching through giant spin-orbit torque in a magnetically doped topological insulator heterostructure. *Nature Mater*. **13**, 699-704 (2014).

14. Kondou, K. *et al*. Fermi-level-dependent charge-to-spin current conversion by Dirac surface states of topological insulators. *Nature Phys*. **12**, 1027-1031 (2016).

15. Lesne, E. *et al*. Highly efficient and tunable spin-to-charge conversion through Rashba coupling at oxide interfaces. *Nature Mater*. **15**, 1261-1266 (2016).

16. Rojas-Sánchez, J. -C. *et al*. Spin to charge conversion at room temperature by

spin pumping into a new type of topological insulator: $\alpha$-Sn films. *Phys. Rev. Lett*. **116**, 096602 (2016).

17. Fan, Y. B. *et al*. Electric-field control of spin-orbit torque in a magnetically doped topological insulator. *Nature Nanotech*. **11**, 352-359 (2016).

18. Liu, R. H. *et al*. Control of current-induced spin-orbit effects in a ferromagnetic heterostructures by electric field. *Phys. Rev. B*. **89**, 220409(R) (2014).

19. Weisheit, M. *et al*. Electric field-induced modification of magnetism in thin-film ferromagnets. *Science* **315**, 349-351 (2007).

20. Fang, D. *et al*. Spin-orbit-driven ferromagnetic resonance. *Nature Nanotech.* **6**, 413-417 (2011).

21. Kurebayashi, H. *et al*. An antidamping spin-orbit torque originating from the Berry curvature. *Nature Nanotech*. **9**, 211-217 (2014).

22. Ciccarelli, C. *et al*. Room-temperature spin-orbit torque in NiMnSb. *Nature Phys*. **12**, 855-860 (2016).

23. Fabian, J. *et al*. Semiconductor spintronics. *Acta Physica Slovaca*. **57**, 565-907 (2007).

24. Nitta, J. *et al*. Gate control of spin-orbit interaction in an inverted $In_{0.53}Ga_{0.47}As/In_{0.52}Al_{0.48}As$ heterostructure. *Phys. Rev. Lett.* **78**, 1335-1338 (1997).

25. Sze, S. M. Semiconductor devices 2$^{nd}$ edn (John Wiely & Sons, Inc, 2002).

26. Liu, C. *et al*. Electrical detection of ferromagnetic resonance in ferromagnet/*n*-GaAs heterostructures by tunneling anisotropic magnetoresistance. *Appl. Phys. Lett.* **105**, 212401 (2014).

27. Liu, C. *et al*. Dynamic detection of electron spin accumulation in ferromagnet-semiconductor devices by ferromagnetic resonance. *Nature Commun.* **7**, 10296

(2016).

28. Jonker, B. T. *et al*. Enhanced carrier lifetimes and suppression of midgap states in GaAs at a magnetic metal interface. *Phys. Rev. Lett.* **79**, 4886-4889 (1997).

29. Qaiumzadeh, A., Duine, R. A., & Titov, M. Spin-orbit torques in two-dimensional Rashba ferromagnets. *Phys. Rev.* B **92,** 014402 (2015).

30. Kato, Y. K. *et al*. Current-induced spin polarization in strained semiconductors. *Phys. Rev. Lett.* **93**, 176601 (2004).

31. Stern, N. P. *et al*. Current-induced polarization and the spin Hall effect at room temperature. *Phys. Rev. Lett.* **97**, 126603 (2006).

32. Engel, H-A. *et al*. Out-of-plane spin polarization from in-plane and magnetic fields. *Phys. Rev. Lett.* **98**, 036602 (2007).

33. Ishihara, J. *et al*. Direct imaging of gate-controlled persistent spin helix state in a modulation-doped GaAs/AlGaAs quantum well. *Appl. Phys. Expre.* **7**, 013001 (2014).

34. Luengo-Kovac, M. *et al*. Gate control of the spin mobility through the modification of the spin-orbit interaction in two-dimensional system. *Phys. Rev. B*. **95**, 245315 (2017).

35. Dettwiler, F. *et al*. Stretchable persistent spin helices in GaAs quantum wells. *Phys. Rev. X* **7**, 031010 (2017).

36. Gambardella, P. *et al*. Current-induced spin-orbit torques. *Phil. Trans. R. Soc. A* **369**, 3175-3197 (2011).

37. Slonczewski, J. C. Current-driven excitation of magnetic multilayers. *J. Mag. Mag. Mater.* **159**, L1-L7 (1996).

38. Berger, L. Emission of spin waves by a magnetic multilayer traversed by a current. *Phys. Rev. B* **54**, 9353-9358 (1996).

**Acknowledgements**

L. Chen thanks M. Kammermeier, M. Buchner and C. Gorini for fruitful discussions. L. Chen is grateful for support from Alexander von Humboldt Foundation. This work is support by the German Science Foundation (DFG) via SFB 689 and SFB 1277.

**Author contribution**

L.C. planned the study. L.C. and R. I. fabricated the devices. L.C. collected and analysed the data. M.K., D.S. and D.B. grew the samples. M.V did the COMSOL simulations. M.G. and J.F. did the first principle calculations and theoretical input. L.C. wrote the manuscript with input from J. F., C.H.B. and D.W.. All authors discussed the results.

**Competing financial interests**

The authors declare no competing financial interests.

**Data availability**

The data that support the plots within this paper and other findings of this study are available from the corresponding author upon reasonable request.

**Figure captions**

**Figure 1 | Electric-field modulation of spin-orbit fields at the Fe/GaAs (001) interface.** Schematic band bending of a Fe/$n$-GaAs Schottky diode under reverse (**a**) and forward (**b**) bias. (**c**) Calculated interfacial electric-field $\mathbf{E}_{in}$ obtained by solving

the one-dimensional Poisson equation for the non-constant doping profile of *n*-GaAs (Methods). The magnitude of interfacial spin-orbit interaction is proportional to $\mathbf{E}_{in}$. Wave-vector **k** dependence of Rashba- (**d**) and Dresselhaus (**e**) spin-orbit fields which dominate at the epitaxial Fe/GaAs interface with $C_{2v}$ symmetry.

**Figure 2 | Electric-field modulation of induced spin-orbit fields measured by spin-orbit ferromagnetic resonance.** (**a**) Schematic of the circuit and device structure for spin-orbit ferromagnetic resonance measurements. A microwave current **j**(*t*), generated by a signal generator, passes through a bias tee into the Fe/GaAs (001) layer, where time dependent interfacial iSOFs **h**(*t*) are generated due to the spin-orbit interaction at the Fe/GaAs interface. **h**(*t*) drives the magnetization dynamics **M**(*t*), and leads to a resistance variation *R*(*t*) due to the anisotropic magneto-resistance of Fe. The coupling between *R*(*t*) and *j*(*t*) produces a rectified d.c. voltage *V*. By measuring *V*, the magnitude of the interfacial iSOFs can be quantified. To modulate the iSOFs by an electric-field, Fe/*n*-GaAs samples have been used, where a Schottky barrier is formed at the interface. Thus, by adding a gate-voltage $V_G$ between Fe and *n*-GaAs, the magnitude of the iSOFs can be controlled. Here, $\varphi_M$ is the magnetization angle with respect to the [100] orientation of GaAs. (**b**) d.c. voltage spectra for several gate-voltages for a [100]-oriented device measured at a microwave frequency of 12.5 GHz and $\varphi_M$ = 225°. The input microwave power is 10 mW. An offset voltage, $V_{offset}$, has been subtracted. (**c**) Typical Lorentzian fit of the voltage spectrum, from which the magnitudes of $V_{sym}$ and $V_{a\text{-}sym}$ are obtained. (**d**) $V_G$ dependence of $V_{a\text{-}sym}$ and $V_{sym}$ for

$\varphi_M$ = 225°. Magnetization angle $\varphi_M$ dependence of $V_{\text{a-sym}}$ (**e**) and $V_{\text{sym}}$ (**f**) under different $V_G$ for a [100]-oriented device. The solid lines in (**e**) and (**f**) are fits to Eqs. 3 and 4, respectively.

**Figure 3 | Gate-voltage dependence of in-plane induced spin-orbit fields.** $V_G$ dependence of in-plane Bychkov-Rashba- (**a**) and Dresselhaus- (**b**) iSOFs obtained from the [100]-oriented device. The solid lines are linear fits, and the modification rate of Rashba-iSOF $m_R^I$ and Dresselhaus-iSOF $m_D^I$ is obtained from the slopes. (**c**) Absolute value of $m_R^I$ and $m_D^I$ for devices along [100] and [010] directions. The magnitude of $|m_R|$ is larger than $|m_D|$ for both orientations. The error bars in (**c**) are standard deviation from the linear fits.

**Figure 4 | Tuning of in-plane induced spin-orbit fields.** Polar plot of in-plane iSOFs. The arrows represent direction and relative strength of $\mathbf{h}_{\text{eff}}$, $\mathbf{h}_{\text{eff}} = \mathbf{h}_R^I + \mathbf{h}_D^I$, along the four main directions for a current density of $1\times10^{10}$ Am$^{-2}$. Note that the magnitude of the iSOFs for [$\bar{1}$10] has been amplified by a factor of 10, and the orange arrows along [110] and [$\bar{1}$10] orientations have been shifted for clarity. The solid lines show the polar plot of the spin-orbit energy splitting, $\Delta\varepsilon_{\text{SO}} = 2\mu_B|\mu_0\mathbf{h}_{\text{eff}}|$ for $V_G = -0.88$ V and +0.07 V. The full angular dependence of $|\mathbf{h}_{\text{eff}}|$ is extrapolated from the magnitude of field-vectors along the four main orientations.

**Figure 5 | Gate-voltage dependence of the out-of-plane induced spin-orbit fields.**

(**a**) Gate-voltage $V_G$ dependence of the out-of-plane induced spin-orbit fields originating from Bychkov-Rashba (**a**) and Dresselhaus (**b**) spin-orbit interaction for the [100]-oriented device. The solid lines in (**a**) and (**b**) are linear fits to the experimental data. From the slope, the modification rate of $h_R^z$ and $h_D^z$ is obtained. (**c**) Absolute values of $m_R^z$ and $m_D^z$ obtained from the linear fits of the out-of-plane iSOFs for devices along [100] and [010] directions. The error bars in (**c**) are standard deviation from the linear fits.

**Supplementary Information**

# Electric-field modification of interfacial spin-orbit field-vector

L. Chen[1*], M. Gmitra[2], M. Vogel[1], R. Islinger[1], M. Kronseder[1], D. Schuh[1], D. Bougeard[1], J. Fabian[2], D. Weiss[1] and C. H. Back[1]

[1]Institute of Experimental and Applied Physics, University of Regensburg, 93040 Regensburg, Germany.

[2]Institute of Theoretical Physics, University of Regensburg, 93040 Regensburg, Germany.

*Corresponding author, e-mail: lin.chen@ur.de

**Table of Contents:**

1. Microwave current distribution in Fe/*n*-GaAs

2. Bias-voltage dependence of resonance field, linewidth and anisotropic magneto-resistance

3. Symmetry of iSOFs with respect to positive and negative magnetic field

4. Electric-field control of iSOFs for another sample

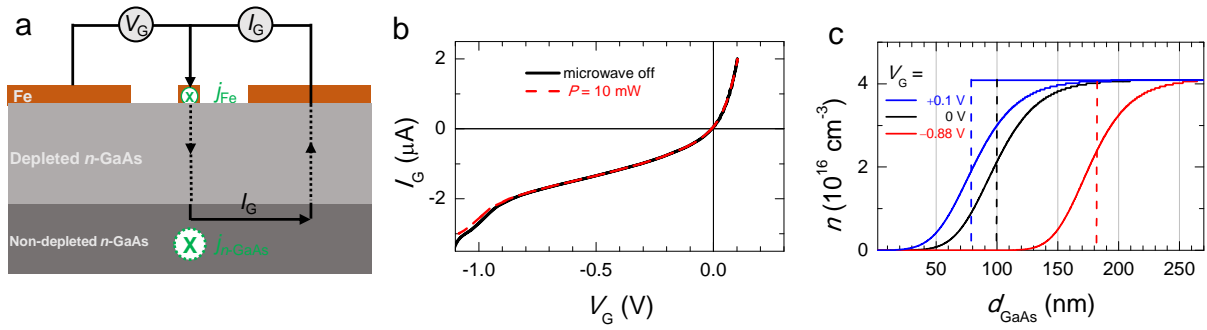

**Figure S1 | $V_G$ - $I_G$ characterization of the Schottky barrier and simulated carrier distribution in $n$-GaAs.** (**a**) Schematic cross section of the device used for Fig. 2a. A microwave current $j_{Fe}$ passes through the Fe stripe and an inductive microwave current is induced in the non-depleted $n$-GaAs. To tune the interfacial spin-orbit interaction, a bias current $I_G$ is applied from the middle Fe stripe to the remote side contact, and the corresponding interfacial gate-voltage $V_G$ is measured between the middle Fe stripe and another remote side contact. (**b**) $V_G$ - $I_G$ characteristics of the Schottky junction with microwave on and off. (**c**) Simulated carrier distribution profile of $n$-GaAs at different $V_G$.

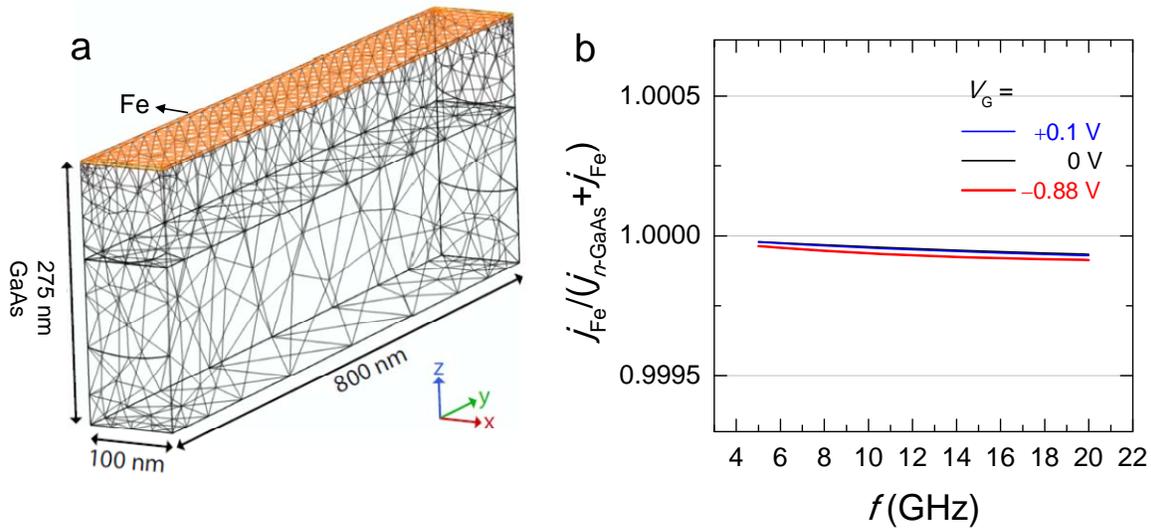

**Figure S2 | Determination of microwave current distribution in Fe/$n$-GaAs using COMSOL 5.3**

(**a**) The current density distribution in Fe/$n$-GaAs stack is calculated by using the COMSOL 5.3 finite elements method software (RF Module). The stack is modeled in 3D using a triangular mesh with a maximum feature size of 3 nm for discretization. The sample size for simulation is scaled down to 8 × 1 μm², and, for clarity, the schematic shown in (**a**) is further reduced down to 0.8 × 0.1 μm². In the simulation, the applied microwave power is 10 μW and the excitation frequency is varied from 5 GHz to 20 GHz. The conductivity of Fe and non-depleted GaAs is set to $3.5 \times 10^6$ S/m and $5.2 \times 10^3$ S/m, respectively. To simplify the calculations, the carrier distribution of GaAs under different $V_G$ is assumed to be a step function distribution as indicated by the dashed lines in Fig. S1c. (**b**) Calculated ratio of the current density in the Fe layer $j_{Fe}$ in relation to the total current density $j_{Fe}+j_{n\text{-GaAs}}$, $j_{Fe}/(j_{Fe}+j_{n\text{-GaAs}})$, as a function of microwave frequency. Here, $j_{n\text{-GaAs}}$ is the inductive microwave current density in un-depleted $n$-GaAs. One can see that most of the current flows in the Fe layer, and the inductive current in non-depleted $n$-GaAs is negligibly small.

## 1. Microwave current distribution in Fe/*n*-GaAs

There could be three kinds of driving forces for magnetization dynamics, which are determined by the microwave current distribution in the Fe layer and the un-depleted *n*-GaAs: one is the interfacial spin-orbit field due to the current flow in Fe, and the other two are the Oersted field $h_{Oe}$ and the Dresselhaus field $h_D^{bulk}$ due to the current flow in the un-depleted *n*-GaAs. Note that the Dresselhaus field in the bulk of *n*-GaAs $h_D^{bulk}$ is essentially different from the interfacial Dresselhaus field at the Fe/GaAs interface discussed in the main text. To minimize $h_{Oe}$ and $h_D^{bulk}$, the sample layout is designed by reducing the thickness (275 nm) and carrier concentration ($4\times10^{16}$ cm$^{-3}$) of *n*-GaAs. Figure S2b shows the COMSOL simulation results[1] of the ratio of the current density in Fe $j_{Fe}$ with respect to the total current density $j_{Fe}+j_{n\text{-GaAs}}$, $j_{Fe}/(j_{Fe}+ j_{n\text{-GaAs}})$, as a function of microwave frequency. Here $j_{n\text{-GaAs}}$ is the inductive current density in un-depleted *n*-GaAs. The magnitude of $j_{Fe}/(j_{Fe}+ j_{n\text{-GaAs}})$ is ~ 1, with a variation far less than 1% in the whole frequency range. This indicates that most of the microwave current flows in the Fe layer, and the inductive current flow in the un-depleted *n*-GaAs is negligibly small.

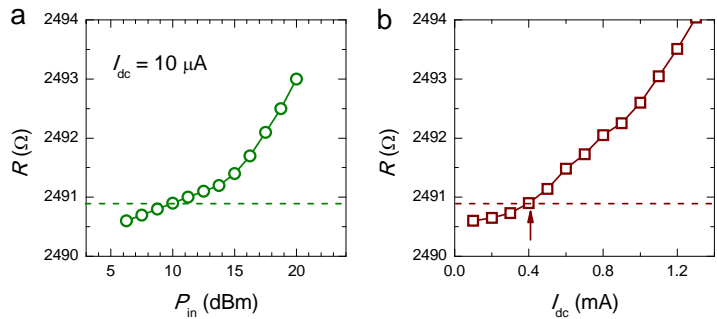

**Figure S3 | Calibration of the microwave current using Joule heating induced resistance change.** (**a**) Resistance of the Fe layer as a function of microwave input power measured at a small d.c. current of 10 µA. (**b**) Resistance of the Fe layer as a function of d.c. current.

The actual microwave current density $j_{Fe}$ is calibrated by the Joule heating induced resistance increase[2]. Figure S3a shows the resistance of the Fe stripe as a function of microwave power with the application of a d.c. current of 10 µA, and the same measurement is done by changing the magnitude of d.c. current as shown in Fig. S3b. By comparing the resistance change in Figs. S3a and b, the microwave current is determined to be 0.4 mA, and the corresponding microwave current density $j_{Fe}$ is $1\times10^{10}$ Am$^{-2}$.

By using $j_{Fe}/j_{n\text{-GaAs}}$ of $2\times10^4$ (Fig. S2b) and $j_{Fe}$ of $1\times10^{10}$ Am$^{-2}$, $j_{n\text{-GaAs}}$ is determined to be $5\times10^5$ Am$^{-2}$. The Oersted field $h_{Oe}$ can be calculated by $\mu_0 h_{Oe} = \mu_0 j_{n\text{-GaAs}} d_{n\text{-GaAs}}/2$, which is 0.05 µT by using $d_{n\text{-GaAs}}$ of 170 nm (Fig. S1c). In comparison with the previous measurement in (Ga,Mn)As[2], the Dresselhaus field $h_D^{bulk}$ in bulk un-depleted GaAs can be estimated to be $10^{-3}$ µT. The magnitude of $h_{Oe}$ and $h_D^{bulk}$ is three orders smaller than $h_R^I$ and $h_D^I$ obtained in the main text (note that, due to the finite spacing between the Fe layer and the un-depleted $n$-GaAs, the effective current-induced fields in $n$-GaAs acting on Fe can be even smaller.). These results indicate that a possible excitation induced by inductive microwave current in the un-depleted $n$-GaAs is negligible, and the main driving force is the spin-orbit field at the Fe/GaAs interface.

The simulation shown in Fig. S2b suggests that the value of $j_{Fe}/(j_{Fe}+ j_{n\text{-GaAs}}) \sim 1$ with variations of the amplitude far less than 1% for different $V_G$. This indicates that the microwave current density in Fe is almost independent of $V_G$. Here, we further exclude this possibility that modification of $h_R$ and $h_D$ is due to modulation of microwave current density as follows:

If the modification of $h_R$ and $h_D$ would be really due to a modulation of the microwave current density $j_{Fe}$, the following form of the current density would be expected: $j_{Fe} = j^0_{Fe}(1+ mV_G)$ with $m$ defining the modification rate of $j_{Fe}$ and $j^0_{Fe}$ is $j_{Fe}$ at $V_G = 0$ V. In this case one can re-write Eq. 3 of the

main text as

$$-\frac{V_{\text{a-sym}}^{[100]} 2M}{\Delta \rho j_{Fe}^{0} l \operatorname{Re}(\chi^{I})} = h^{I} = (1+mV_{G})\left(-h_{D}^{I} \sin\phi_{M} + h_{R}^{I} \cos\phi_{M}\right)\sin 2\phi_{M}. \quad (S1)$$

Here $h_R^I$ and $h_D^I$ are the in-plane Rashba and Dresselhaus iSOF at 0 V with $|h_R^I| \sim |h_D^I|$ (see Figs. 3a and b in the main text). Therefore, the modification rate of Rashba-iSOF $m_R^I$ and Dresselhaus-iSOF $m_D^I$ due to the modulation of $j_{Fe}$ would be

$$|m_R^I|=|mh_R^I|,\ |m_D^I|=|mh_D^I|, \quad (S2)$$

Since $|h_R^I| \sim |h_D^I|$, we should have $|m_R^I| \sim |m_D^I|$. However, our experimental results show that $|m_R^I|$ is ~ 5 times larger than $|m_D^I|$, being at odds with the above assumption. Therefore, we exclude the possibility that the modification of $h_R$ and $h_D$ is due to a modulation of the microwave current density $j_{Fe}$.

## 2. Bias-voltage dependence of resonance field, linewidth and anisotropic magneto-resistance

Figure S4 shows the gate-voltage $V_G$ dependence of the resonance field $H_R$ and the linewidth $\Delta H$. One can see that both $H_R$ and $\Delta H$ are independent of $V_G$, indicating that the magnetic properties of Fe, e.g, the magnitude of $M$, $\operatorname{Re}(\chi^I)$ and $\operatorname{Im}(\chi_a^O)$, are not modulated by the electric-field. This can be reasonably understood as the follows: the capacitance $C$ of a Schottky contact[3] is estimated to be ~ $10^{-7}$ Fcm$^{-2}$, and the modulation of the interfacial electron density can be expressed by $\Delta n = (C/e)V_G$, where $e$ is the electronic charge. Under a gate-voltage of $-1$ V, $|\Delta n|$ is determined to be ~ $10^{12}$ cm$^{-2}$, which is at least three orders of magnitude smaller than the sheet carrier concentration of bulk Fe. We

also confirm in Fig. S5 that the magnitude of the anisotropic resistance is independent of gate-voltage.

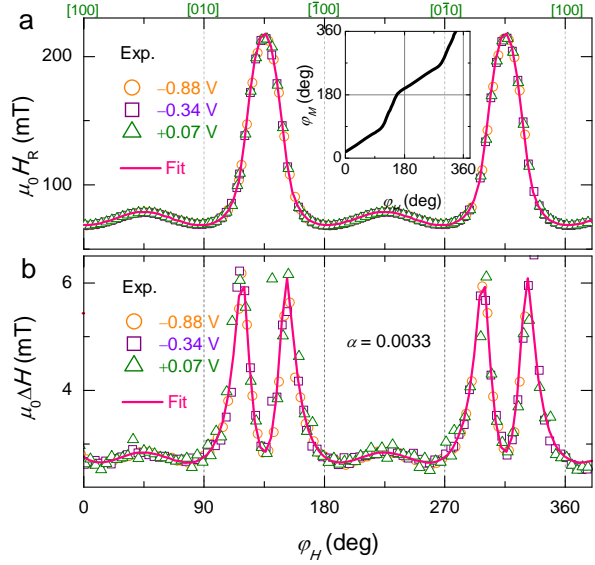

**Figure S4 | Gate-voltage dependence of resonance field and linewidth.** Magnetic-field angle $\varphi_H$ dependence of resonance field $H_R$ (**a**) and linewidth $\Delta H$ (**b**) under different gate-voltages for the [100]-orientated device. By fitting the $\varphi_H$ dependence of $H_R$, one can obtain the magnitude of the effective perpendicular magnetic anisotropy field $\mu_0 H_K$ of 1506.5 mT, the biaxial magnetic anisotropy field $\mu_0 H_B$ of 34.0 mT, the uniaxial magnetic anisotropy field $\mu_0 H_U$ of 72.0 mT and the Landé $g$ factor of 2.06. The inset of (**a**) shows $\varphi_H$ as a function of magnetization angle $\varphi_M$. By fitting the $\varphi_H$ dependence of $\Delta H$, the damping constant is obtained to be 0.0033.

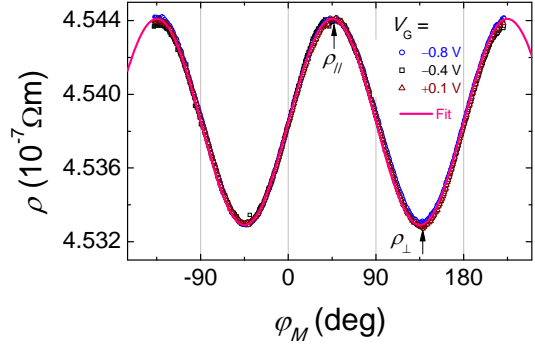

**Figure S5 | Gate-voltage dependence of the anisotropic magneto-resistance.** $\varphi_M$ dependence of the resistivity of the [110]-orientated device, which is gate-voltage $V_G$ independent. In the measurement, an external magnetic field of 1 T has been used, which leads to $\varphi_M = \varphi_H$. The solid line is a fit, from which the magnitude of anisotropic magneto-resistance, $\Delta\rho = \rho_{//} - \rho_{\perp}$, is determined to be $1.1 \times 10^{-9}$ Ωm.

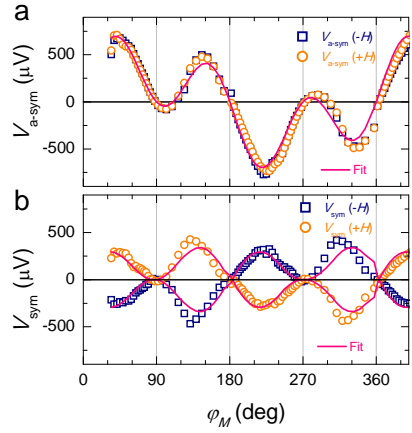

**Figure S6 | Symmetry of d.c. voltage with respect to positive and negative magnetic-field.** (**a**) Magnetization angle $\varphi_M$ dependence of $V_{\text{a-sym}}(-H)$ and $V_{\text{a-sym}}(+H)$ for a [100]-orientated device, which shows $V_{\text{a-sym}}(-H) = V_{\text{a-sym}}(+H)$. (**b**) $\varphi_M$ dependence of $V_{\text{sym}}(-H)$ and $V_{\text{sym}}(+H)$, which shows $V_{\text{sym}}(-H) = -V_{\text{sym}}(+H)$. The solid lines in (**a**) and (**b**) are fits according to Eqs. (3) and (4), respectively. The thickness of the sample is 5 nm, which is measured at a frequency of 11 GHz with a microwave power of 168 mW.

## 3. Symmetry of iSOFs with respect to positive and negative magnetic field

The magnitude of $V_{\text{a-sym}}(-H)$ and $V_{\text{a-sym}}(+H)$ as a function of magnetization angle $\varphi_M$ is shown in Figure S6a, one can see the symmetry of $V_{\text{a-sym}}(-H) = V_{\text{a-sym}}(+H)$; while for $V_{\text{sym}}$, shown in Figure S6b, one can see that $V_{\text{sym}}(-H) = -V_{\text{sym}}(+H)$ holds. This is because $V_{\text{a-sym}}$ is induced by in-plane iSOFs $h^{\text{I}}$ (see Eq. 3 in the main text), which only depends on the current direction with symmetry $h^{\text{I}}(-H) = h^{\text{I}}(+H)$; on the other hand $V_{\text{sym}}$ is induced by out-of-plane SOFs $h^{\text{O}}$, being anti-dampling like and having the symmetry $h^{\text{O}}(-H) = -h^{\text{O}}(+H)$ (see Fig. 4 of Ref. 4).

## 4. Electric-field control of iSOFs for another sample

We have measured another sample consisting the following layers: Fe (1.3 nm)/$n^+$-GaAs (10 nm, $5 \times 10^{17}$ cm$^{-3}$)/$n^+ \rightarrow n$-GaAs (15 nm)/$n$-GaAs (250 nm, $4 \times 10^{16}$ cm$^{-3}$). Here the doping level of $n^+$-GaAs has been reduced by a factor of 2 to increase the width of the Schottky barrier. As shown in Fig. S7a, the interfacial resistance $R_{\text{in}}$ of this sample is ~ 8 times larger than the sample presented in the main text (Fig. S1b). The gate-voltage dependences of iSOFs are shown from Fig. S7b to Fig. S7e and summarized in Table S1. One can see that the modification rate of each iSOF becomes smaller in comparison with the higher $n^+$-doped sample. The smaller modification rate could result from the smaller change of $E_{\text{in}}$ with $V_G$, being due to the increased width of the Schottky barrier.

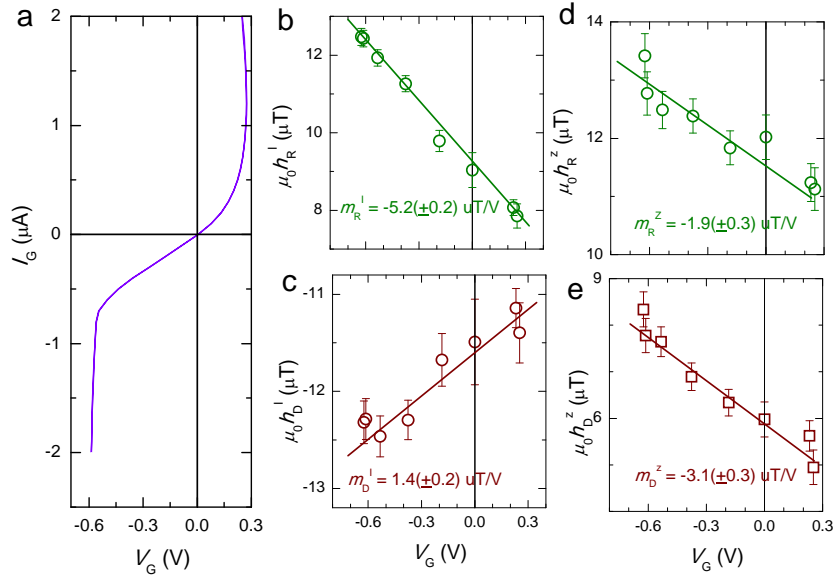

**Figure S7 | Electric-field control of iSOFs for another sample consisting of the following layers: Fe (1.3 nm)/$n^+$-GaAs (10 nm, $5\times10^{17}$ cm$^{-3}$)/$n^+\rightarrow n$-GaAs (15 nm)/$n$-GaAs (250 nm, $4\times10^{16}$ cm$^{-3}$)** (**a**) $V_G$ - $I_G$ characterization of the Schottky barrier for the device orientated along [100]. The device dimension is 20 μm × 50 μm. $V_G$ dependence of in-plane Bychkov-Rashba (**b**), in-plane Dresselhaus (**c**) out-of-plane Bychkov-Rashba (**d**) and out-of-plane Dresselhaus (**e**) iSOF.

| $t_{Fe}$ (nm) | $n^+$(cm$^{-3}$) | $R_{in}$ (kΩ/μm$^2$) | $|m_R^I|$ (μT/V) | $|m_D^I|$ (μT/V) | $|m_R^Z|$ (μT/V) | $|m_D^Z|$ (μT/V) |
|---|---|---|---|---|---|---|
| 4.0 | $1\times10^{18}$ | 0.19@−0.5 V | 10.7±0.5 | 2.0±0.2 | 8.4±0.9 | 4.3±1.0 |
| 1.3 | $5\times10^{17}$ | 0.86@−0.5 V | 5.2±0.2 | 1.4±0.2 | 1.9±0.3 | 3.1±0.3 |

**Table S1. Comparison of electric-field effect on samples with different Fe thickness and $n+$ doping levels.**

# References

1. COMSOL rf package, www.comsol.com

2. Fang, D. *et al.* Spin-orbit-driven ferromagnetic resonance. *Nature Nanotech*. **6**, 413-417 (2011).

3. Sze, S. M. Semiconductor Devices 2nd edn (John Wiley & Sons, Inc, 2002).

4. Chen, L. *et al.* Robust spin-orbit torque and spin-galvanic effect at the Fe/GaAs (001) interface at room temperature. *Nature Commun.* **7**, 13802 (2016).